\documentclass[aps,preprint,showpacs]{revtex4}
\usepackage{graphicx,epsfig}
\usepackage{amssymb}
\begin{document}

\title{Scattering and absorption sections by an improved Schwarzschild black hole}

\author{Omar Pedraza$^1$}
\email{omarp@uaeh.edu.mx}
\author{L. A. L\'opez$^1$}
\email{lalopez@uaeh.edu.mx (corresponding-author)}
\author{L. O. T\'ellez Tovar}
\email{luis.osvaldo.501@gmail.com}

\affiliation{$^1$ \'Area Acad\'emica de Matem\'aticas y F\'isica, UAEH, 
Carretera Pachuca-Tulancingo Km. 4.5, C. P. 42184, Mineral de la Reforma, Hidalgo, M\'exico.}

\begin{abstract}

In this contribution, we investigate the scattering and absorption sections of the improved Schwarzschild black hole. The differential scattering section is analysed using three complementary approaches: the classical approximation, the semi-classical approximation, and the partial wave technique.  We show that, while the classical scattering section exhibits only small deviations from the standard Schwarzschild case, the semi-classical and partial wave analyses reveal differences in the interference pattern and in the amplitude. Also, the absorption section is computed using the partial wave method and compared with the sinc approximation. We find that both approaches present deviations that appear in the low–frequency regime, where the partial wave result approaches the horizon area. Our results indicate that quantum corrections in the Schwarzschild can lead to modifications in scattering and absorption properties, providing further insight into the phenomenology of quantum corrected black holes.
\end{abstract}


\maketitle

\section{Introduction}

General Relativity (GR), proposed by Einstein predicts the existence of black holes (BHs), whose simplest solution is the Schwarzschild metric \cite{Schwarzschild:1916uq}. This solution describes a static and spherically symmetric space-time generated by a compact object characterized only by its mass. From a semi-classical perspective, black holes emit Hawking radiation, with a temperature inversely proportional to their mass. As a consequence, the mass of the BH decreases as the evaporation process progresses. However, it remains unclear whether this process continues until the complete evaporation of the black hole or whether it stops when the system reaches scales close to the Plank regime, where quantum gravitational effects are expected to become important.

In order to incorporate quantum corrections into the classical description of BHs, several approaches have been proposed. One of them is based on the Wilsonian re-normalization group applied to gravity, which leads to the so-called improved Schwarzschild BH \cite{Bonanno:2000ep}. In this scenario, the classical metric is modified through scale-dependent couplings, introducing quantum corrections that become relevant at small scales. The properties of this geometry have been investigated in different contexts, including geodesic motion \cite{Mandal:2022stf} and quasi-normal modes \cite{Rincon:2020iwy}, revealing interesting deviations from the standard Schwarzschild BH.

As a complementary aspect of geodesic equations and quasi-normal mode analyses, one can also investigate the  interaction of fields around the one BH. In particular, the study of scattering and absorption processes provides important information about the interaction between fields and the gravitational potential of the BH. The scattering section encodes how incoming waves are deflected by the space-time curvature, while the absorption section measures the fraction of radiation captured by the horizon. These quantities are directly related to several astrophysical  phenomena, such as the black hole shadows, and the accretion of matter and radiation.

In the high frequency regime, the propagation of waves can be approximated by null geodesics, allowing the classical scattering section to be obtained from the deflection of trajectories \cite{Collins:1973xf}. On the other hand, semi-classical methods, such as the glory approximation \cite{PhysRevD.31.1869}, take into account interference effects produced by partial waves with different angular momenta. The partial wave method \cite{PhysRevD.22.2331} provides a full wave analysis by reducing the problem to a Schrödinger-like equation that describes the radial propagation of scalar fields around the black hole.

In addition to scattering, the absorption properties of black holes play a fundamental role in understanding how energy and matter interact with the event horizon. The absorption section in the limit for high frequencies (sinc approximation) can be written as the sum of the called geometric section and the oscillatory part of the absorption section \cite{Decanini:2011xi}. Also the absorption section can be obtained with partial wave method.

Motivated by these considerations, in this work we analyse the scattering and absorption sections of massless scalar waves in the background of an improved Schwarzschild black hole. 

The paper is organized as follows. In Sec.~II we present the improved Schwarzschild black hole solution and analyse its horizon structure. In Sec.~III we discuss the classical and semi-classical descriptions of the scattering section. Also we compute the scalar scattering section using the partial wave approach. Section~IV is devoted to the analysis of the absorption section and its comparison with the sinc approximation. Finally, in Sec.~V we summarize our main results.

\section{Improved Schwarzschild black hole}

The Schwarzschild solution describes the gravitational field of a static and spherically symmetric mass and arises as a vacuum solution of Einstein's field equations. In the renormalization group (RG) approach to quantum gravity, the gravitational coupling becomes scale dependent due to quantum effects. Following the RG improvement procedure proposed by Bonanno and Reuter \cite{Bonanno:2000ep}, Newton's constant is promoted to a running coupling $G(k)$, where $k$ represents the RG scale. By identifying this scale with a characteristic distance of the space-time, one obtains an effective coupling $G(r)$. Replacing the classical constant $G$ by $G(r)$ in the Schwarzschild lapse function leads to the improved Schwarzschild BH, which incorporates quantum corrections at short distances while recovering the classical behaviour at large scales.

\begin{figure}[!h]
	\centering
	\includegraphics[scale=0.85]{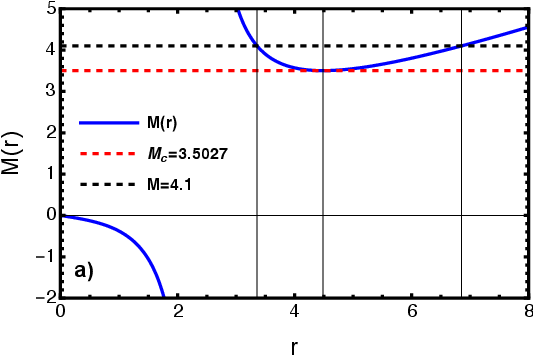}
	\includegraphics[scale=0.883]{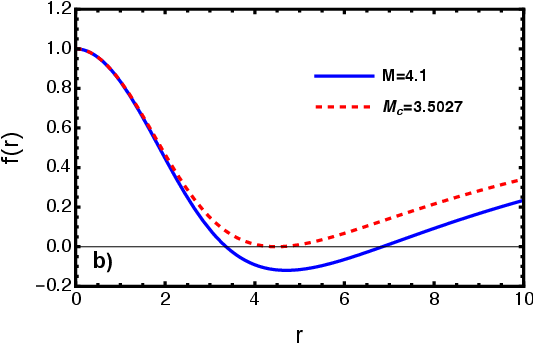}
	\caption{ a) The blue line in the plot on the left represents the evolution of the mass as a function of the radius $r$; the red dotted and black dotted lines are the critical mass $M_c = 3.50274$ and $M = 4.1$ respectively. The critical value $r_c=4.4841$ and the horizon radii $r_{in}=3.3584$ and $r_{out}=6.8497$ are also shown. b) The plot shows the evolution of $f(r)$ where we can see that the number of event horizons depends on the mass value.}
	\label{f1}
\end{figure}

The line element of improved Schwarzschild BH  \cite{Rincon:2020iwy} is given by;

\begin{equation}\label{mfa}
ds^2=-f(r)dt^2+\frac{dr^2}{f(r)}+r^2d\theta^2+r^2\sin^2\theta d\phi^2\,,
\end{equation}
where
\begin{equation}
f(r)=1-\frac{2Mr^2}{r^3+\xi\left(r+\gamma M\right)}\,
\end{equation}
Here, $M$ is the parameter of mass, $\xi=\frac{118}{15\pi}$ and $\gamma=\frac{9}{2}$ represents the quantum parameters. The number of horizons depends entirely on the choice of the value of parameter $M$. The horizons of improved Schwarzschild BH are determined by the positive roots of expression $f(r)=0$, this condition leads to the following equation (\ref{ec.his});
\begin{equation}\label{ec.his}
r^3-2Mr^2+\xi r+\xi\gamma M=0\,,
\end{equation}
Thus, for determinate the range of values of $M$ that represents a BH solution, we use the method applied in \cite{Pedraza:2020uuy}. From (\ref{ec.his}), the mass parameter as function of $r$ is given by
\begin{equation}
M(r)=\frac{r\left(r^2+\xi\right)}{2r^2-\gamma\xi}.
\end{equation}
Here, we can see that $M\to 0$ when $r\to0$, and for $r\to\infty$, $M\to\infty$. Besides, the function $M(r)$ is discontinuous at $r=\sqrt{\frac{\gamma\xi}{2}}$. The behavior of the $M(r)$ is showed in Fig.~\ref{f1}. We see that there is a critical value $M_c = 3.50274$ for the mass of the improved Schwarzschild BH with $r_c = 4.4841$, which leads to two well-defined cases: first, for values of $M>M_c$ we have a BH with two horizons, while for a value of $M=M_c$ there is a single horizon that corresponds to $r_c$ as we can see in Fig.~\ref{f1} for the particular value $M = 4.1$.

\section{Scattering sections}

In this section, we compute the scattering section of the improved Schwarzschild BH using three complementary approaches: the classical geodesic approximation, the semi-classical glory approximation, and the partial wave method. This combined analysis allows us to investigate how quantum corrections modify the scattering behaviour in comparison with the standard Schwarzschild BH.

\subsection{Classical Scattering}

A first approximation to obtain the scattering  section consists in analysing geodesics propagating from infinity toward the black hole. This approach relies on the fact that, in the high-frequency limit, incident waves can be effectively described as propagating along null geodesics \cite{Collins:1973xf}.

The motion of test particles along null geodesics is governed by the Lagrangian $\mathcal{L}=- \frac{1}{2}\dot{x}^{\mu}\dot{x}_{\mu}=0$, where $\dot{x}^{\mu}=dx^{\mu}/d\tau$, and $\tau$ denotes the affine parameter along the geodesic.

As the improved Schwarzschild BH (\ref{mfa}) is static and spherically symmetric, the coordinates $t$ and $\phi$ are cyclic. Consequently,  the energy $E$ and the angular momentum $L$ are conserved quantities defined as; 

\begin{equation}
E = \left(1-\frac{2Mr^2}{r^3+\xi\left(r+\gamma M\right)}\right)\dot{t}, \qquad  L = r^{2}\sin \theta \dot{\phi}
\end{equation}

If we only consider the motion in the plane $\theta=\pi / 2$. Solving for $\dot{r}^2$, we obtain $\dot{r}^2=E^2-V_{eff}$, with;

\begin{equation}\label{Vef}
V_{eff}= \left(1-\frac{2Mr^2}{r^3+\xi\left(r+\gamma M\right)}\right)\frac{L^2}{r^2}.
\end{equation}

The conserved quantities allow the particle trajectories to be characterized in terms of the impact parameter $b=L/E$. In particular, the introduction of the critical impact parameter $b_c$ is of special relevance, as it separates trajectories that are scattered back to infinity from those that are captured by the BH, for any $ b> b_c$ there is scattering. 

In order to obtain the critical impact parameter, we first consider the radial equation where the effective potential (\ref{Vef}) is defined. By introducing the change of variable $u=1/r$, the radial equation can be written as: 

\begin{equation}\label{Ang}
	\left(\frac{du}{d\phi}\right)^{2}=\frac{1}{b^{2}}-f(1/u)u^{2}\,.
\end{equation}

Differentiating (\ref{Ang}) with respect to $\phi$, we obtain

\begin{equation}\label{Ang1}
	\frac{d^{2}u}{d\phi^{2}}=-\frac{u^{2}}{2}\frac{df(1/u)}{du}-uf(1/u)\,.
\end{equation}

Then if we impose the condition $\frac{d^{2}u}{d\phi^{2}}=0$, the positive root corresponds to the radius of the critical orbit for null geodesics, denoted by $u_c$. Equivalently, the radius of the circular null geodesic $r_c$ can be obtained from the conditions $V_{\rm eff}'(r_c)=0$ and $V_{\rm eff}(r_c)=E^{2}$. Substituting $u_c$ into Eq.~(\ref{Ang}), we obtain the critical impact parameter $b_c$

In the case of geodesics coming from the infinity to a turning point $u_{0}$, the deflection angle $\theta\left( b\right)$ is given by;

\begin{equation}\label{impact parameter}
	\theta\left( b\right)=2\phi\left( b\right)-\pi\,,
\end{equation}
where
\begin{equation}\label{Integral}
	\phi=\int_0^{u_0}du \left(\frac{1}{b^2}-u^2 f(1/u)\right)^{-1/2}\,.
\end{equation}

We have a relation between the impact parameter $b(\theta)$ and the classical scattering section given by;

\begin{equation}\label{Sec}
	\frac{d\sigma}{d\Omega}=\frac{1}{\sin \theta} \sum  b(\theta) \left|\frac{d b(\theta)}{d \theta}\right|\,.
\end{equation}

where the sum accounts for the possibility that null geodesics may orbit the BH several times. As a consequence, a geodesic can circle the BH multiple times before eventually escaping to infinity (see \cite{Dolan:2006vj} for details). 

Now for a given value of the mass parameter and impact parameters satisfying $b>b_c$, the turning point $u_0$ can be obtained from (\ref{Ang}). We then numerically integrate the elliptic equation (\ref{Integral}) and invert (\ref{impact parameter}) to determine $b(\theta)$. Substituting this result into (\ref{Sec}), we compute the classical scattering section.

\begin{figure}[!h]
	\centering
	\includegraphics[scale=0.9]{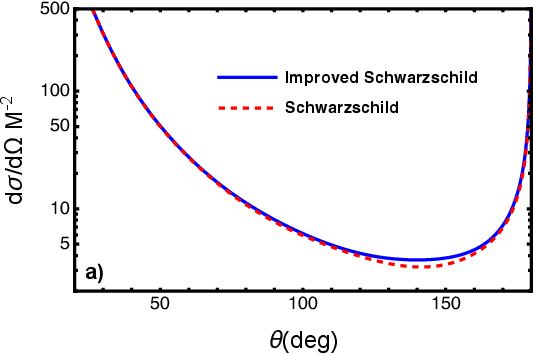}
	\includegraphics[scale=0.88]{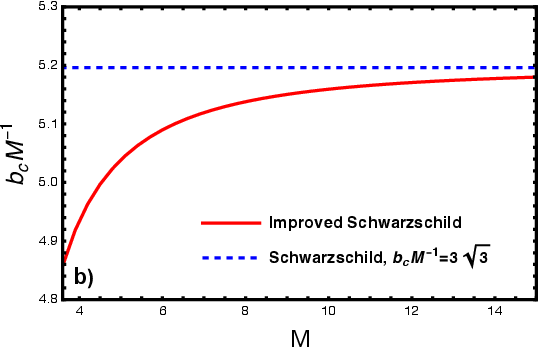}
	\caption{  a) Comparison of the classical scattering cross section for the improved Schwarzschild BH and the Schwarzschild BH with $M=4$.  b) Behaviour of the normalized critical impact parameter $b_{c} M^{-1}$ for the improved Schwarzschild BH}
	\label{f2}
\end{figure}

Fig.~\ref{f2} a) displays the classical scattering section obtained from (\ref{Sec}). We observe that, at large scattering angles, the improved Schwarzschild black hole produces a larger section than the standard Schwarzschild case. This behaviour indicates that the quantum corrections modify the effective gravitational potential in such a way that null geodesics experience a stronger deflection in the near-horizon region. This is reflected in the relation  $\left(\frac{d\sigma}{d\Omega}\right)_{ImpSchw}>\left(\frac{d\sigma}{d\Omega}\right)_{Schw}$.

Also Fig.~\ref{f2} b) displays the behaviour of the normalized critical impact parameter $b_c M^{-1}$ for the improved Schwarzschild BH. The improved solution exhibits a clear dependence on the mass parameter. For masses close to the critical value, $b_c M^{-1}$ is reduced, indicating that quantum corrections modify the structure of unstable null geodesics and, consequently the effective capture region. As $M$ increases, the ratio $b_c M^{-1}$ recovers the classical behaviour, then quantum effects become relevant for small masses and directly impact observable quantities.

\subsection{Semi--classical Scattering}

As a second approach to the analysis of scattering sections, we now turn to the semi-classical approximation, which incorporates interference effects. In particular, the classical scattering section in Eq.~(\ref{Sec}) does not account for the interference between partial waves with different angular momenta. To include these effects, we employ the semi-classical approach, commonly referred to as the glory scattering approximation~\cite{PhysRevD.31.1869}. This method is especially suitable in the high-frequency regime ($\omega \gg 1$), where scalar plane waves can be treated in terms of null geodesics, and interference arises from rays scattered near the backward direction. Within this framework, the scattering section for spherically symmetric black holes is given by:

\begin{equation}
	\frac{d\sigma_g}{d\Omega}=2\pi \omega  b_g^2 \left| \frac{d b}{d\theta}\right|_{\theta=\pi}J_{2s}^2(\omega  b_g \sin\theta)\,,
\end{equation}

with $\omega$ as the wave frequency, $J_{2s}^2$ are the Bessel function of first kind of order $2s$ where $s$ represents the spin, in particular, $s=0$ for scalar waves; and $b_g$ denotes the impact parameter of the reflected waves ($\theta \sim\pi$). This semi-classical approximation is valid for $M\omega \gg 1$ where $M$ is the BH mass (the approximation can still reproduce the  scattering  for $\omega M \sim 1$.). From the glory approximation, it is known that the interference fringe widths depend inversely on the impact parameter $b_{g}$. 

\begin{figure}[!h]
	\centering
	\includegraphics[scale=0.89]{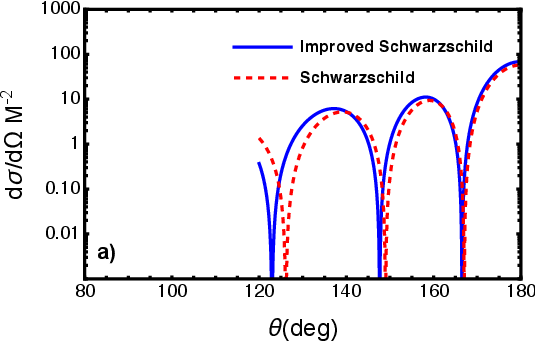}
	\includegraphics[scale=0.89]{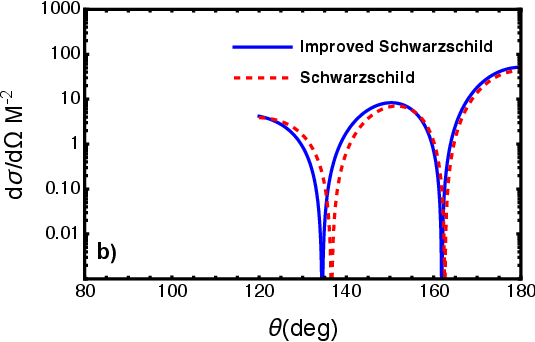}
	\caption{ The graphs show the behavior of semi-classical scattering section for improved Schwarzschild and Schwarzschild BHs with $M=4$. In the figure a),  we use $M\omega=2$, while in figure b) we use $M\omega=1.5$.}
	\label{f3}
\end{figure}

The semi-classical scattering sections for the improved Schwarzschild and Schwarzschild BHs are compared in Fig.~\ref{f3}. It can be observed that there are differences between the two semi-classical sections, which depend on the values of the black hole mass $M$ and the wave frequency $\omega$. These differences are reflected in the widths of the interference fringes, as shown in Fig.~\ref{f3}.

\subsection{Partial Waves Approach}

As a final step in the study of scattering sections, it is necessary to consider the wave description in terms of a massless scalar field. In this framework, the scattered waves are modeled by solutions of the Klein--Gordon equation. From a physical perspective, the partial wave method provides a complete wave description of the scattering process, incorporating both diffraction and interference effects.

Accordingly, in this subsection we investigate the scattering of massless scalar waves by the improved Schwarzschild black hole, which is governed by the Klein--Gordon equation:

\begin{equation}\label{Gordon}
\square \Phi=g^{\mu \nu}\nabla_{\mu} \nabla_{\nu} \Phi= \frac{1}{\sqrt{g}}\partial_{\mu}(\sqrt{-g}g^{\mu \nu}\partial_{\nu}\Phi)=0\,.
\end{equation}

We shall be interested in monochromatic plane waves, so we consider the ansat;

\begin{equation}\label{field}
\Phi_{\omega}=\sum_{lm}\frac{\hat{\phi_l}(r)}{r}Y^{m}_{l}(\theta, \phi)e^{-i\omega t}\,,
\end{equation}	

where $Y^{m}_{l}(\theta, \phi)$ are the scalar spherical harmonics, $l$ denotes the angular degree, and $\omega$ represents the frequency. Introducing the radial partial wave functions $\Phi_{\omega}$ in (\ref{Gordon}) and doing separation of variables we obtained 

\begin{equation}\label{radial}
\frac{d^{2}\hat{\phi_l}(r)}{dr^{2}_{*}}+\omega^{2}\hat{\phi_l}(r) - V_{l}(r)\hat{\phi_l}(r) =0\,.
\end{equation}

This equation, known as the Regge--Wheeler equation, is written in terms of the tortoise coordinate $r_{*}=r_{*}(r)$, defined for $r \in I$ by the relation $(dr_{*}/dr) = (1/f(r))$. On the other hand, $V_{l}$ denotes the Regge--Wheeler potential associated with the massless scalar field, which is given by;

\begin{equation}\label{Vee}
V_{l}(r)=f(r)\left(\frac{l(l+1)}{r^{2}}+\frac{f^{'}(r)}{r}\right)\,.
\end{equation}

The scalar potential (\ref{Vee}) vanishes in both asymptotic limits  $r_{*}\rightarrow -\infty$ (near the horizon) and $r_{*}\rightarrow \infty$ (spatial infinity). It is well known that suitable boundary conditions must be imposed to solve (\ref{radial}). For an asymptotically flat spacetime, such as the improved Schwarzschild black hole, the boundary conditions are given by:

\begin{equation}\label{ec.sphip}
\hat{\phi_l}(r_{*})= \left\{ \begin{array}{lr}
e^{-i\omega r_{*}}+R_{\omega l}e^{i\omega r_{*}}, & r_{*} \to+\infty\\
T_{\omega l}e^{-i\omega r_*}, & r_* \to-\infty
\end{array} \right.
\end{equation}

where the $\left|R_{\omega l}\right|^2 $ and $\left|T_{\omega l}\right|^2 $ are the reflection and transmission coefficients, respectively and the flux conservation implies that $\left|R_{\omega l}\right|^2 +\left|T_{\omega l}\right|^2=1$.  

Now the scalar differential scattering section, in a static and spherically symmetric four dimensional space-time, is given in terms of partial waves by:

\begin{equation}
	\frac{d\sigma}{d\Omega}=\arrowvert f(\theta)\arrowvert^{2}\,,
\end{equation}

where $f(\theta)$ is the scattering amplitude given by;

\begin{equation}\label{functionpartial}
	f(\theta)=\frac{1}{2i \omega}\sum_{l=0}^{\infty}(2l+1)\left[\left(-1\right)^{l+1}R_{\omega l}-1\right]P_{l}(\cos \theta).
\end{equation}

However as noted in \cite{PhysRev.95.500} \cite{Dolan:2006vj} \cite{Li:2025lvl}, the series in (\ref{functionpartial}) converges poorly, making its direct summation difficult. This arises from the need for infinitely many Legendre polynomials to describe the divergence at $\theta = 0$. So circumvent this issue, Eq.~(\ref{functionpartial}) can be rewritten in the form:

\begin{equation}
    2i\omega f(\theta) = \sum_{l=0}a_l^{(0)}P_{l}(\cos \theta),
\end{equation}
we can define the $m$th reduced series
\begin{equation}
    (1 - \cos\theta)^{\mathrm{m}} 2i\omega f(\theta) = \sum_{l=0}^{\infty}a_l^{(\mathrm{m})}P_{l}(\cos \theta)
\end{equation}
where the coefficients are given by the iterative formula
\begin{equation}
    \label{iterative}
    a_l^{(i+1)} = a_l^{(i)} - \frac{l+1}{2l+3}a_{l+1}^{(i)} - \frac{l}{2l-1}a_{l-1}^{(i)}.
\end{equation}
With
\begin{equation}
    a_l^{(0)} = (2l+1)\left[\left(-1\right)^{l+1}R_{\omega l}-1\right].
\end{equation}
In the other hand, we can also obtain the absorption cross section given by
\begin{equation}
    \label{absparciales}
    \sigma_{abs} = \sum_{l=0}^\infty \frac{\pi}{\omega^2}(2l + 1) T_{\omega l}.
\end{equation}

The scattering cross section in the partial-wave method is obtained by numerically solving the radial equation (\ref{radial}) subject to the boundary conditions (\ref{ec.sphip}). We impose the near-horizon condition at $r = r_h + \delta$, with $\delta \ll 1$, while the outer boundary (numerical infinity) is chosen depending on the value of $l$, in our case, we set $r = 500$ for $l = 40$. To improve convergence, we adopt $a^{2}$.

\begin{figure}[!h]
	\centering
	\includegraphics[scale=0.9]{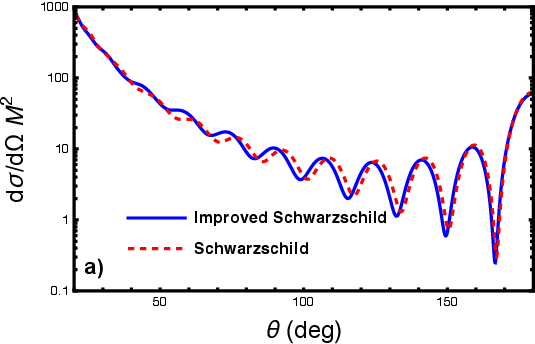}
	\includegraphics[scale=0.9]{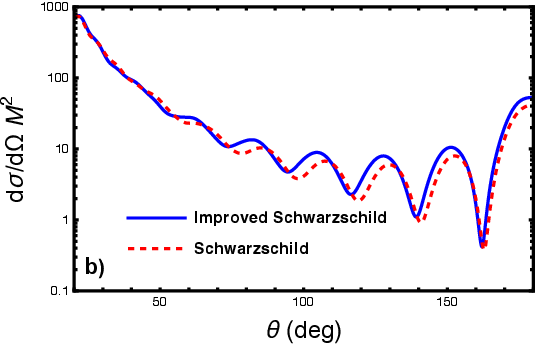}
	\caption{  Behavior of scalar differential scattering section of improved Schwarzschild BH with $M=4$ and Schwarzschild BHs. In the figure a),  we use $M\omega=2$, while in the figure b) $M\omega=1.5$.}
	\label{f4}
\end{figure}

In Fig. \ref{f4} we compare the scattering section for the improved Schwarzschild BH, as well as for the Schwarzschild BH with different frequency values. We can see that when $M\omega = 2$ there is no significant difference between the two BHs. However, for $M\omega = 1.5$ (see Fig. \ref{f4} b)), we can see that the amplitude and width of the interference fringes of the scattering section are greater for the improved Schwarszschild BH compared to the original.

\begin{figure}[!h]
	\centering
	\includegraphics[scale=0.9]{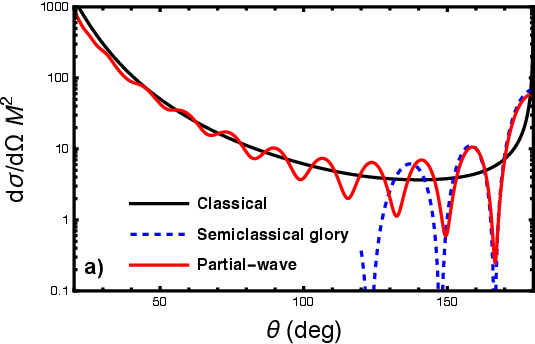}
	\includegraphics[scale=0.9]{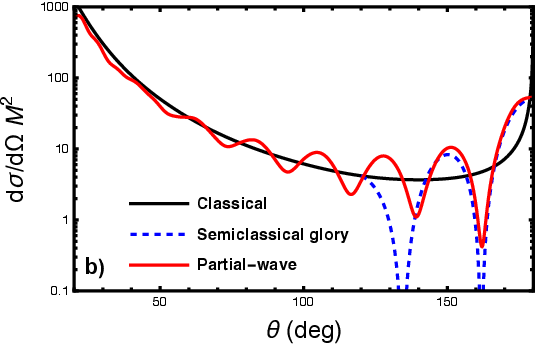}
	\caption{Comparison of partial wave, semi-classical and classical  approaches for the  scattering  section, for $M = 4$ with $M\omega = 2$ (figure a)) and  $M\omega = 1.5$ ( figure b)) for the improved Schwarzschild BH.}
	\label{f6}
\end{figure}

In Fig.~\ref{f6}, we present a comparison between the partial wave method, the geodesic and glory approximations for the improved Schwarzschild. We find that the glory approximation provides an excellent agreement with the partial wave results for angles $\theta \lesssim \pi$, while the geodesic approximation accurately describes the small-angle regime.

\section{Absorption Cross Section}

In addition to scattering section, the analysis of absorption section offers valuable insights into black hole physics, as it is directly connected to the accretion of matter and fields, as well as to the formation and characteristics of black hole shadows.

In the low-frequency regime, where the wavelength is large than the black hole size, the absorption is governed by the horizon geometry and the section reduces to the horizon area~\cite{Higuchi:2001si} \cite{Das:1996we}.

In contrast, in the high-frequency limit, wave propagation is well approximated by null geodesics, and the absorption section approaches the classical capture section, $\sigma_{\mathrm{geo}} = \pi b_c^{2}$, determined by the critical impact parameter (see subsection A). 

Moreover, in the eikonal regime, the absorption section exhibits oscillations around $\sigma_{\mathrm{geo}}$, arising from interference effects of waves orbiting near the photon sphere. These oscillations can be expressed in terms of the parameters of unstable null geodesics~\cite{PhysRevD.83.044032} as;

\begin{equation}
    \sigma_{osc} = -4 \pi \frac{\lambda b_c^2}{\omega} e^{-\pi \lambda b_c} \sin \left(2\pi \omega b_{c}\right),
\end{equation}
where $\lambda$ is the Lyapunov exponent \cite{Cardoso:2008bp}:
\begin{equation}
    \lambda^2 = \frac{f(r_c)}{2r_c^2}\left[2f(r_c) - r_c^2f''(r_c)\right],
\end{equation}
with $r_c$ the radius of the unstable null circular orbits; again, $b_c$ is the critical impact parameter. 

Finally, the absorption cross section, in the high frequencies limit, is proportional to $\sigma_{geo}$ and $\sigma_{osc}$:
\begin{equation}
    \label{sinc}
    \sigma_{sinc} \approx \sigma_{geo} + \sigma_{osc}.
\end{equation}

Eq. (\ref{sinc}) is known as the sinc approximation.

Returning to the partial wave method, the absorption section can be obtained from (\ref{absparciales}) (see subsection~C). These results can then be compared with the sinc approximation (\ref{sinc}), in order to extract further information about the absorption section of the improved Schwarzschild BH.

\begin{figure}[!h]
	\centering
	\includegraphics[scale=0.89]{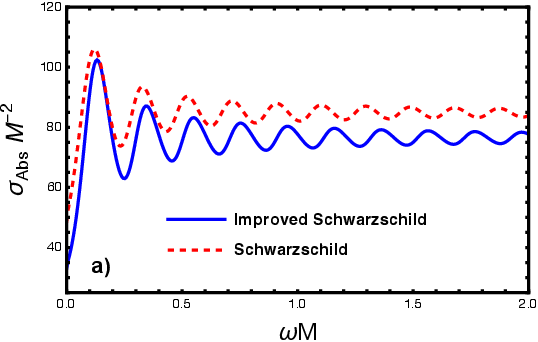}
	\includegraphics[scale=0.89]{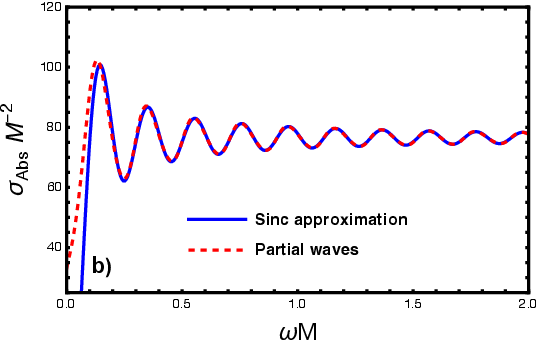}
	\caption{In a), we compare the absorption cross section, obtained through the partial wave method, for the improved Schwarzschild BH and Schwarzschild BH with $M=4$. In b) the absorption cross section is contrasted with the sinc approximation for improved Schwarzschild BH with $M=4$}
	\label{fabs}
\end{figure}

In Fig.~\ref{fabs} a), we compare the absorption section obtained via the partial wave approach (\ref{absparciales}), for the improved Schwarzschild  BH and Schwarzschild BH. We observe that, in the low-frequency limit, the absorption cross section approaches the black hole horizon area. In the other hand, in Fig. ~\ref{fabs} b) we show the sinc approximation, it can be observed that the absorption cross section obtained using the sinc approximation shows good agreement for larger values of $M \omega$.

The amplitude of the absorption cross section tends to a constant value as $\omega M$ increases (see Fig.~\ref{fabs} a) and  b)). For a fixed value of $M$, we observe that the absorption section of the Schwarzschild BH is larger than of  improved  Schwarzschild BH. This behavior can be understood from the modification of the null geodesic structure, which affects the radius of the photon sphere and the associated critical impact parameter $b_c$. As a consequence, the geometric cross section $\sigma_{\mathrm{geo}} \propto b_c^{2}$ increases, leading to a higher absorption at high frequencies.

\section{Conclusions}

In this work, we have analysed the scattering and absorption sections of a massless scalar field in an improved Schwarzschild black hole, employing three complementary approaches: the classical geodesic approximation, the semi-classical glory approximation, and the full partial-wave method.

While the classical scattering section exhibits only small deviations from the Schwarzschild black hole, effects arise at the semi-classical and wave levels. In particular, the interference patterns are modified, with their structure depending sensitively on both the black hole mass and the wave frequency. These effects can be directly traced to changes in the properties of unstable null geodesics, which are encoded in the critical impact parameter and the photon sphere.

Regarding absorption, we find that in the high-frequency regime, the agreement between the sinc approximation and the partial-wave results confirms the dominant role of null geodesics in governing wave propagation. In contrast, deviations at low frequencies highlight the importance of wave effects and the role of the horizon geometry.

Overall, our analysis demonstrates that scattering and absorption  processes provide a suitable framework to investigate how quantum corrections modify the gravitational properties of the Schwarzschild black hole. The modifications observed here suggest that such effects could, in principle, leave imprints on observable quantities related to black hole optics, such as shadows, lensing, and wave signatures.

\section*{ACKNOWLEDGMENT}

The authors acknowledge to SECIHTI-SNII, Mexico.

\bibliographystyle{unsrt}

\bibliography{bibliografia}

\end{document}